\documentclass[12pt]{article}

\usepackage{amsfonts}
\usepackage{epsfig}
\usepackage{latexsym}
\usepackage{graphicx}
\def\bequ{\begin{equation}}
\def\eequ{\end{equation}}
\def\barr{\begin{array}}
\def\earr{\end{array}}
\def\half{{1\over 2}}
\def\ben{\begin{equation}}
\def\een{\end{equation}}
\def\bena{\begin{eqnarray}}
\def\eena{\end{eqnarray}}

\def\spa#1{\phantom{\fbox{\rule[-#1cm]{0cm}{0cm}}}}

\def\drawbox#1#2{\hrule height#2pt
        \hbox{\vrule width#2pt height#1pt \kern#1pt
              \vrule width#2pt}
              \hrule height#2pt}

\def\Fund#1#2{\vcenter{\vbox{\drawbox{#1}{#2}}}}
\def\Asym#1#2{\vcenter{\vbox{\drawbox{#1}{#2}
              \kern-#2pt       
              \drawbox{#1}{#2}}}}

\def\funda{\Fund{6.5}{0.4}}

\def\symm{\funda\kern-0.4pt\funda}

\def\b1{e^0}

\newcommand{\be}{\begin{equation}}
\newcommand{\ee}{\end{equation}}
\def\bea{\begin{eqnarray}}
\def\eea{\end{eqnarray}}
\def\Tr{\mbox{Tr}}

\def\del {\partial}
\def\nn{\nonumber}

\def\half {{1 \over 2}}

\def\be{\begin{equation}}
\def\ee{\end{equation}}
\def\bea{\begin{eqnarray}}
\def\eea{\end{eqnarray}}

\def\lesssim{\mathrel{\hbox{\rlap{\hbox{\lower4pt\hbox{$\sim$}}}\hbox{$<$}}}}
\def\gtrsim{\mathrel{\hbox{\rlap{\hbox{\lower4pt\hbox{$\sim$}}}\hbox{$>$}}}}

\begin{document}

\hfuzz=100pt
\title{{\Large \bf{Models with Quartic Potential of Dynamical SUSY Breaking in Meta-Stable Vacua}}}
\author{\\Shinji Hirano\footnote{hirano@nbi.dk}
  \spa{0.5} \\
{{\it The Niels Bohr Institute}}
\\ {{\it Blegdamsvej 17, DK-2100 Copenhagen}}
\\ {{\it Denmark}}}
\date{}

\maketitle
\centerline{}

\begin{abstract}
We search for models of dynamical SUSY breaking in meta-stable vacua which might have dual string descriptions with a few brane probes.  
Two models with quartic superpotential are proposed: One of them might be closely related to the dual gauge theory to the flavored Maldacena-Nu\~nez geometry by  Casero, Nu\~nez, and Paredes with a few additional brane probes corresponding to massive flavors. 
The other model might be dual to the Klebanov-Strassler geometry with one fractional D3-brane and a few D7-branes as probes.
\end{abstract}

\section{Introduction}

The supersymmetry (SUSY) offers irresistible elegance to various branches in theoretical high energy physics, yet the nature has not hinted its existence to date.
So the SUSY, if it exists, has to be broken in some way at the higher energy scale.
The dynamical SUSY breaking models thus far constructed tend to be baroque, which compromises the elegance of the SUSY. To spare the virtue of supersymmetry, it would be desirable to find the way to dynamically break SUSY in a minimal manner. 

Recently Intriligator, Seiberg, and Shih (ISS) found a remarkably simple and beautiful mechanism of dynamical SUSY breaking, once one accepts the SUSY breaking in meta-stable vacua \cite{Intriligator:2006dd,Intriligator:2007cp}.  
The ISS model is just ${\cal N}=1$ $SU(N_c)$ SQCD with $N_f$ light quarks in the range $N_c+1\le N_f<{3\over 2}N_f$. 
The simplicity of their mechanism opened up a new direction for the phenomenologically viable model-buildings of dynamical SUSY breaking (see \cite{Intriligator:2007cp} for the references for the recent developments).

Besides phenomenological applications, it would be of great interest to see if a similar mechanism is realizable in the realm of ${\cal N}=1$ gauge/string duality. This is the theme of the present paper.
Models of meta-stable dynamical SUSY breaking in this context were previously proposed in \cite{Argurio:2006ew,Argurio:2006ny}. Also somewhat related to this theme is the realization of the ISS-type models via brane configurations in Type IIA \cite{Ooguri:2006bg,Franco:2006ht,Giveon:2007fk}, MQCD \cite{Bena:2006rg}, and noncritical string theory \cite{Murthy:2007qm}, and by geometric engineerings \cite{Ooguri:2006pj,Aganagic:2006ex,Tatar:2006dm}.\footnote{There is overwhelmingly a long list of literatures which might be related to this theme or otherwise very interesting and worth being mentioned. Some of the works can be found in the references of \cite{Intriligator:2007cp}. We apologize for unintended omissions. Any suggestions for updating and correcting the references will be appreciated.}

The non-singular gravity/string duals to ${\cal N}=1$ gauge theories are the Klebanov-Strassler (KS) geometry \cite{Klebanov:2000hb}, its generalization by Butti, Gra\~na, Minasian, Petrini, and Zaffaroni \cite{Butti:2004pk}, the Maldacena-Nu\~nez (MN) geometry \cite{Maldacena:2000yy}, and the flavored MN geometry by Casero, Nu\~nez, and Paredes  \cite{Casero:2006pt}.
In this paper we search for the models of dynamical SUSY breaking in meta-stable vacua which might be dual to the KS and the flavored MN geometries with a small number of probe branes. 
The ISS-type models typically require the number of flavors to be greater than the number of colors. In the gauge/gravity duality, the number of colors must be taken to be large. 
So the most naive idea of adding flavors as probes would not work.
A large number of massless flavors, however, are built-in manifestly in the flavored MN geometry and as bi-fundamentals of the $A_2$ quiver in the KS geometry.  
The ISS model with massless quarks was studied by Franco and Uranga \cite{Franco:2006es}. In order to break SUSY, the total number of massless and massive quarks must be greater than the number of colors, but the massive quarks alone can be just a few. There is a subtlety concerning the meta-stability of the SUSY breaking vacua due to the modulus associated with the massless quarks. However, this modulus can be lifted by a quartic potential for the massless quarks \cite{Amariti:2006vk}.
Given this observation, roughly speaking, the idea is to add a few massive flavors as probes in the KS and flavored MN geometries and that the quartic potential to lift the would-be troublesome modulus is present in the gauge theories dual to these geometries.\footnote{The validity of the probe approximation for flavor branes must be argued with caution. No matter how small the number of flavor branes is, they generate the logarithmic gravitational potential in the UV. So as we go sufficiently far in the UV, their back-reaction always prevails. However, we are interested in the IR dynamics. There the probe approximation may be justified. We thank Aki Hashimoto for his stressing this point.}

If our claim for the KS geometry is solidified, it would provide a very natural way of breaking SUSY in the GKP-KKLT flux compactification \cite{Giddings:2001yu}\cite{Kachru:2003aw} in the same spirit as \cite{Kachru:2002gs,Argurio:2006ny}.

The organization of our paper is as follows. 
In section 2 and 3 we review the ISS model and the ISS model with massless quarks respectively. 
In section 4 we add the quartic potential for the massless quarks to the model reviewed in section 3 and discuss its virtues. This is claimed to be closely related to the dual gauge theory to the flavored MN geometry. In section 5, as an application of previous sections, we consider the KS gauge theory with massive quarks, and discuss under what conditions the dynamical SUSY breaking occurs in meta-stable vacua. In section 6 we discuss how the SUSY breaking model in section 5 might be realized in the KS geometry, and close the discussions with a comment on its application to the KKLT model.

\section{The ISS model}

The ${\cal N}=1$ $SU(N_c)$ SQCD with $N_f$ quarks in the range $N_c<N_f<3N_f$ has $N_c$ SUSY vacua. 
However, Intriligator, Seiberg, and Shih \cite{Intriligator:2006dd} found a finer structure of vacua deeper in the IR. When the quarks are massive but very light and the theory is in the free magnetic phase $N_c+1 < N_f < {3\over 2}N_c$ or in the confining phase $N_f=N_c+1$ \cite{Seiberg:1994bz,Seiberg:1994pq,Intriligator:1995au}, there exist meta-stable SUSY breaking vacua farther down in the IR.

To study the IR physics for $N_c+1 < N_f < {3\over 2}N_c$, the Seiberg duality \cite{Seiberg:1994pq} was used: The dual magnetic theory is IR free and enables us to study more details of the dynamics deep in the IR. 
The magnetic theory is $SU(N_f-N_c)$ SQCD with $N_f$ dual quarks $q_c^{\mbox{ }i}$ and $\widetilde{q}_i^{\mbox{ }c}$, $N_f^2$ singlets (mesons) $M_i^{\mbox{ }j}$, and the superpotential $W={1\over\Lambda}\Tr\, qM\widetilde{q}+\Tr\, mM$, where $m_i^{\mbox{ }j}$ is the quark mass matrix, $c=1,\cdots, N_f-N_c$, and $i,j=1,\cdots,N_f$.\footnote{The dynamical scales of the dual pair are related by $\Lambda^{3N_c-N_f}\widetilde{\Lambda}^{3(N_f-N_c)-N_f}=(-1)^{N_f-N_c}\hat{\Lambda}^{N_f}$, where $\Lambda$ is the scale of the electric theory, $\widetilde{\Lambda}$ that of the magnetic theory, and $\hat{\Lambda}$ the scale entering due to the ambiguity associated with the rescaling of the dual quarks \cite{Intriligator:1995au}. We have chosen $\hat{\Lambda}=\Lambda$ for our convenience.}
Since the magnetic theory is IR free, the K\"ahler potential is smooth and takes the form 
$K={1\over\beta}\Tr\left(q^{\dagger}q+\widetilde{q}^{\dagger}\widetilde{q}\right)+{1\over\alpha|\Lambda|^2}\Tr\, M^{\dagger}M+\cdots$, where the ellipses denote the higher order terms which are negligible in the IR. The real positive dimensionless coefficients $\alpha$ and $\beta$ are of order one. Their precise values are not known, but we are only concerned about the qualitative results. 

In order to illustrate the main results of ISS, we only consider the equal mass case $m=-{\mu^2\over\Lambda}{\bf 1}_{N_f}$. By rescaling $q$, $\widetilde{q}$, and $\Phi\equiv{M\over\Lambda}$ appropriately, we have the canonically normalized K\"ahler potential and the superpotential
\be
W=h\Tr\,q\Phi\widetilde{q}-h\mu^2\Tr\Phi\ ,
\label{superpot}
\ee
where $h$ is a dimensionless coupling (marginally irrelevant). The superpotential is not renormalized in all orders in perturbation theory except for the renormalization of the couplings due to the wavefunction renormalizations.

The first point to note is that the supersymmetry is spontaneously broken (in all orders in perturbation theory). This is because the F-term 
\be
F_{\Phi_{i}^{\mbox{ }j}}={\del W\over\del\Phi_{i}^{\mbox{ }j}}
=h\left(\widetilde{q}_{j}^{\mbox{ }c}q_c^{\mbox{ }i}-\mu^2\delta_j^{\mbox{ }i}\right)
\ee
cannot be vanishing, since the rank of the matrix $\widetilde{q}_{j}^{\mbox{ }c}q_c^{\mbox{ }i}$ is at most $N\equiv N_f-N_c<N_f$ while that of $\mu^2\delta_j^{\mbox{ }i}$ is $N_f$ -- the rank condition.
The classical vacua, the minima of the scalar potential, are at 
\bea
\Phi=\left(
\begin{array}{cc}
{\bf 0}_N & {\bf 0}_{N\times (N_f-N)} \\
{\bf 0}_{(N_f-N)\times N} & \Phi_0
\end{array}
\right),
q=\left(
\begin{array}{cc}
q_0\ , & {\bf 0}_{N\times (N_f-N)}
\end{array}
\right),
\widetilde{q}^T=\left(
\begin{array}{cc}
\widetilde{q}_0\ , & {\bf 0}_{N\times (N_f-N)} 
\end{array}
\right),
\eea
where $q_0=\mu e^{\theta}{\bf 1}_N$ and $\widetilde{q}_0=\mu e^{-\theta}{\bf 1}_N$. 
The D-terms vanish on these vacua, and the vacuum energy density of the SUSY breaking vacua is $V_{min}=\left(N_f-N\right)\left|h^2\mu^4\right|$.

The second point concerns the perturbative stability of the SUSY breaking vacua. Let us consider the maximally symmetric vacua, $\Phi_0=0$ and $\theta=0$. Most of the fluctuations about this vacua are massive. However, there remain two kinds of massless modes -- (1) Nambu-Goldstone (NG) modes due to the breaking of the global symmetries\footnote {The magnetic dual theory with the superpotential (\ref{superpot}) has the global symmetries $SU(N_f)_D\times U(1)_B\times U(1)_R$ where $SU(N_f)_D\subset SU(N_f)\times SU(N_f)$. The maximally symmetric vauum breaks the gauge symmetry $SU(N)$ completely, but the global symmetries are broken to $SU(N)_D\times SU(N_f-N)\times U(1)_{B'}\times U(1)_R$ where $SU(N)_D\subset SU(N_f)\times SU(N_f)$.} and (2) the classical moduli $\delta\Phi_0$ and $\mu\mbox{Re}\,\delta\theta$. The former remain massless quantum mechanically, being protected by the symmetries. 
The latter, however, turned out to be lifted at one-loop, acquiring the masses $m_{\Phi_0}^2={\ln 4-1\over 8\pi^2}N|h^4\mu^2|$ and $m_{\theta}^2={\ln 4-1\over 8\pi^2}(N_f-N)|h^4\mu^2|$.
The higher-loops are negligible since the couplings are marginally irrelevant.  
Hence the maximally symmetric vacua 
\be
\Phi_0=0\ ,\quad q_0=\widetilde{q}_0=\mu{\bf 1}_N
\ee
are the perturbatively stable quantum vacua (whose moduli space is parameterized by the NG bosons).

The third point to note is the existence of the SUSY vacua due to the non-perturbative effect, rendering the SUSY breaking vacua meta-stable. 
The SUSY vacua appear elsewhere in the field space of $\Phi$: Consider the physics at the energy scale $E<h\Phi$. Through the cubic coupling the dual quarks $q$ and $\widetilde{q}$ acquire the effective masses of order $|h\Phi|$. Thus they can be integrated out in the low energy effective theory at the energy scale $E$. Now the low energy effective theory becomes pure ${\cal N}=1$ $SU(N)$ Yang-Mills theory. The superpotential of ${\cal N}=1$ $SU(N)$ Yang-Mills is generated by the gaugino condensation and given by $W_{YM}=N\Lambda_L^3$ where $\Lambda_L$ is the dynamical scale of the low energy theory. The matching of the couplings at the energy scale $\left(\det(h\Phi)\right)^{1/N_f}$ yields $\Lambda_L^{3N}=\Lambda^{3N-N_f}\det(h\Phi)$. Thus the superpotential of the low energy effective theory is given by
\be
W_L=W_{YM}-h\mu^2\Tr\Phi=N\left(h^{N_f}\Lambda^{3N-N_f}\det\Phi\right)^{{1\over N}}-h\mu^2\Tr\Phi\ .
\ee
This yields $N_c$ SUSY vacua
\be
\langle h\Phi\rangle=\Lambda\epsilon^{2N/(N_f-N)}{\bf 1}_{N_f}
=\mu\epsilon^{-(N_f-3N)/(N_f-N)}{\bf 1}_{N_f}\ ,
\ee
where $\epsilon=\mu/\Lambda$. 

In order to ensure the validity of the analysis, the parameter $|\epsilon|\ll 1$: The energy scales of the SUSY breaking, the SUSY vacua, and the Landau pole are well separated as
\be
|\mu|\ll|\langle h\Phi\rangle|\ll|\Lambda|\ .
\ee
These inequalities vindicate the use of the magnetic dual description to extract the IR physics. The first inequality in particular justifies integrating out the dual quarks, and implies the longevity of the meta-stable SUSY breaking vacua.

In the confining case $N_f=N_c+1$, there is no magnetic dual description. However, the non-perturbative superpotential is known in terms of the baryons $B^i$ and $\widetilde{B}_i$ and the mesons $M_i^{\mbox{ }j}$ where $i,j=1,\cdots, N_f$. The result is essentially the extrapolation of the $N_c+1<N_f$ case to $N(=N_f-N_c)=1$ with $q=B/\Lambda^{N_c-1}$, 
$\widetilde{q}=\widetilde{B}/\Lambda^{N_c-1}$, and $\Phi=M/\Lambda$, up to the rescalings by numerical constants:
\be
W={1\over\Lambda^{2N_c-1}}\left(\Tr\,BM\widetilde{B}-\det\, M\right)+\Tr\, mM\ ,
\label{confsuppot}
\ee
with the K\"ahler potential $K={1\over\beta|\Lambda|^{2N_c-2}}\Tr\left(B^{\dagger}B+\widetilde{B}^{\dagger}\widetilde{B}\right)+{1\over\alpha|\Lambda|^2}\Tr\, M^{\dagger}M$.
Therefore the same conclusion as the $N_c+1<N_f$ case holds.

In summary, ${\cal N}=1$ SQCD with $N_f$ light flavors in the range $N_c+1\le N_f<{3\over 2}N_c$ has meta-stable SUSY breaking vacua in the deeper IR in addition to $N_c$ SUSY vacua.

\section{The ISS model plus massless quarks}

The models to be proposed which dynamically breaks SUSY in meta-stable vacua and may have dual string descriptions typically contain a large number of massless quarks and a few massive quarks. 
So as a preliminary we first review the ISS model plus massless quarks studied by Franco and Uranga \cite{Franco:2006es}.

Let $N_{f,0}$ and $N_{f,1}$ be the number of massless and massive quarks respectively. 
For the $SU(N_{f,0}+N_{f,1}-N_c)$ dual magnetic theory to be IR free, the theory has to be in the range $N_c+1<N_f=N_{f,0}+N_{f,1}<{3\over 2}N_c$. 
The classical superpotential of the magnetic theory is given by
\be
W=h\left(\Tr\, q_0\Phi_{00}\widetilde{q}_0 
+\Tr\, q_0\Phi_{01}\widetilde{q}_1
+\Tr\, q_1\Phi_{10}\widetilde{q}_0\right)
+\left(h\Tr\, q_1\Phi_{11}\widetilde{q}_1
-h\mu^2 \Tr\, \Phi_{11}\right)\ ,
\label{SUPOTISSplusmassless}
\ee
where $q$s and $\widetilde{q}$s are the dual quarks, and $\Phi$s are the mesons as in the previous section. The subscripts $0$ and $1$ denote the degrees of freedom associated with massless and massive quarks respectively. We have chosen the mass matrix to be $m=-{\mu^2\over\Lambda}{\bf 1}_{N_{f,1}}$.

Exactly as in the ISS model, the supersymmetry is broken in all orders in perturbation theory, if $N\equiv N_{f,0}+N_{f,1}-N_c<N_{f,1}$, {\it i.e.}, $N_{f,0}<N_c$, 
since the F-term 
\be
F_{(\Phi_{11})_{i}^{\mbox{ }j}}={\del W\over\del(\Phi_{11})_{i}^{\mbox{ }j}}
=h\left((\widetilde{q}_1)_{j}^{\mbox{ }c}(q_1)_c^{\mbox{ }i}-\mu^2\delta_j^{\mbox{ }i}\right)
\ee
cannot be vanishing because of the rank condition --  the rank of the matrix $(\widetilde{q}_1)_{j}^{\mbox{ }c}(q_1)_c^{\mbox{ }i}$ is at most $N<N_{f,1}$ while that of $\mu^2\delta_j^{\mbox{ }i}$ is $N_{f,1}$, where $c=1,\cdots, N$ and $i,j=1,\cdots, N_{f,1}$.
Classically the SUSY breaking vacua are at 
\bea
&&q_0=\widetilde{q}_0^T=0\ ,\hspace{.5cm} \Phi_{01}=({\bf 0}_{N_{f,0}\times N}\ , \hat{\Phi}_{01})
\ ,\hspace{.5cm} \Phi_{10}=\left(
\begin{array}{c}
{\bf 0}_{N\times N_{f,0}} \\ 
\hat{\Phi}_{10}
\end{array}
\right)\nn\\
&& q_1=\left(q\ , {\bf 0}_{N\times (N_{f,1}-N)}\right)\ ,\hspace{.5cm}
\widetilde{q}_1^T=\left(\widetilde{q}\ , {\bf 0}_{N\times (N_{f,1}-N)}\right)\ ,\\
&&\Phi_{11}=\left(
\begin{array}{cc}
{\bf 0}_N & {\bf 0}_{N\times (N_{f,1}-N)} \\
{\bf 0}_{(N_{f,1}-N)\times N} & \hat{\Phi}_{11}
\end{array}
\right)\ ,\hspace{.5cm}
\Phi_{00}=\mbox{arbitrary}\ ,\nn
\eea
where $\hat{\Phi}_{01}$ and $\hat{\Phi}_{10}^T$ are $N_{f,0}\times (N_{f,1}-N)$ matrices, 
$q=\mu e^{\theta}{\bf 1}_N$, and $\widetilde{q}=\mu e^{-\theta}{\bf 1}_N$.
The D-terms vanish on these vacua, and the vacuum energy density of the SUSY breaking vacua is $V_{min}=\left(N_{f,1}-N\right)\left|h^2\mu^4\right|$.

Unlike the ISS model, the stability of the SUSY breaking vacua is more subtle and potentially non-perturbatively unstale: Let us consider the maximally symmetric vacua, $\hat{\Phi}_{11}=0$, $\theta=0$, $\hat{\Phi}_{01}=\hat{\Phi}_{10}^T=0$, and $\Phi_{00}=0$. There are again two kinds of massless modes about this vacua -- (1) NG modes and (2) the classical moduli $\delta\hat{\Phi}_{11}$, $\mu\mbox{Re}\,\theta$, $\delta\hat{\Phi}_{01}$, $\delta\hat{\Phi}_{10}$, and $\delta\Phi_{00}$.
The former remain massless quantum mechanically, being protected by the symmetries. 
The latter, except for $\delta\Phi_{00}$, are lifted at one-loop, acquiring the masses 
$m_{\hat{\Phi}_{11}}^2={\ln 4-1\over 8\pi^2}N|h^4\mu^2|$, $m_{\theta}^2={\ln 4-1\over 8\pi^2}(N_f-N)|h^4\mu^2|$, and $m_{\hat{\Phi}_{01}}^2=m_{\hat{\Phi}_{10}}^2={\ln 4-1\over 8\pi^2}N|h^4\mu^2|$.

The modulus $\delta\Phi_{00}$ remains massless at one-loop. It might or might not be lifted at higher-loops. If not, the SUSY breaking vacua is non-perturbatively unstable: Consider the energy scale $E<h\Phi_{00}, h\Phi_{11}$. Since the dual quarks have the effective masses of order $|h\Phi_{00}|$, $|h\Phi_{11}|$, they can be integrated out at this energy scale. Then the low energy effective theory becomes pure ${\cal N}=1$ $SU(N)$ Yang-Mills theory, generating the non-perturbative superpotential by the gaugino condensation $W_{YM}=N\Lambda_L^3$ with $\Lambda_L$ being the dynamical scale of the low energy effective theory. The matching of the gauge couplings at the energy scale 
$\left(\det\, h\Phi_{00}\det\, h\Phi_{11}\right)^{{1\over N_f}}$ reads 
$\Lambda_L^{3N}=\Lambda^{3N-N_f}\det\, h\Phi_{00}\det\, h\Phi_{11}$. Thus the low energy  effective superpotential yields
\be
W_L=N\left(h^{N_f}\Lambda^{3N-N_f}\det\,\Phi_{00}
\det\,\Phi_{11}\right)^{{1\over N}}
-h\mu^2\Tr\,\Phi_{11}\ .\label{NPplusmasslessPot}
\ee
Integrating out $\Phi_{11}$ then gives
\be
W_L=-(N_{f,1}-N)\left({\mu^{2N_{f,1}}\Lambda^{N_f-3N}}\over h^{N_{f,0}}\det\,\Phi_{00}\right)^{{1\over N_{f,1}-N}}\ .
\label{masslessNPpot}
\ee
This leads to a run-away potential. Although the magnetic dual analysis can be trusted only at $|h\Phi_{00}|, |h\Phi_{11}|\ll |\Lambda|$, the run-away behavior at higher energy scale is completed by the electric theory analysis.\footnote{The superpotential (\ref{masslessNPpot}) is the Affleck-Dine-Seiberg potential for ${\cal N}=1$ $SU(N_c)$ SQCD with $N_{f,0}<N_c$ massless quarks. This can be obtained in the electric theory after integrating out $N_{f,1}$ massive quarks at the energy scale $E<|\mu^2/\Lambda|$. For the run-away potential to be completed in the UV, it would be necessary that $|\mu|\gg|\Lambda|$ in this preliminary model. }
Hence if the modulus $\delta\Phi_{00}$ is not lifted at higher-loops, the SUSY breaking vacua are non-perturbatively unstable along the $\Phi_{00}$ direction.

In order to evade this subtlety, Franco and Uranga introduced new $N_{f,0}^2$ singlets  $\left(\Sigma_0\right)_i^{\mbox{ }j}$ ($i,j=1,\cdots, N_{f,0}$) and the additional superpotential
$W_{add}=\mu_0\Tr\,\Sigma_0\Phi_{00}$ which in terms of the electric theory takes the form $W_{add}={\mu_0\over\Lambda}\Tr\, Q_0\Sigma_0\widetilde{Q}_0$. Then $\Phi_{00}$ is fixed to be zero by the equation of motion w.r.t. $\Sigma_0$ and simply is not a modulus already at the classical level.
The SUSY breaking vacua are now meta-stable and the SUSY vacua are at $|\Sigma_0|\to\infty$ where $|\Phi_{00}|\to 0$ and $|\Phi_{11}|\to 0$.\footnote{This endangers the validity of integrating out the quarks. However, as long as the tail of $|\Sigma_0|$ is amputated, no matter how far it is, $|h\Phi_{00}|$ and $|h\Phi_{11}|$ can be made parametrically larger than $|\mu|$, if $|\epsilon|=|\mu/\Lambda|\ll 1$ and $|\mu_0/\mu|\gg 1$. The model in the next section does not have this subtlety.}

\section{A model with quartic superpotential}

We now consider another way of lifting the modulus $\delta\Phi_{00}$ motivated by ${\cal N}=1$ gauge/string duals, in particular, the flavored MN geometry by Casero, Nu\~nez, and Paredes \cite{Casero:2006pt} and the KS geometry \cite{Klebanov:2000hb}. 
One characteristic of the gauge theory duals of these geometries is the presence of the quartic superpotential for the massless flavors.

We thus add the quartic superpotential 
$W_{add}=\lambda (Q_0)_{ci}(\widetilde{Q}_0)^{id}(Q_0)_{dj}(\widetilde{Q}_0)^{jc}$ ($c=1,\cdots, N_c$ and $i,j=1,\cdots,N_{f,0}$) for the massless quarks. In the dual magnetic description the corresponding superpotential is $W_{add}=\widetilde{\lambda}\Tr\,\Phi_{00}^2$ with $\widetilde{\lambda}\propto\lambda\Lambda^2$ \cite{Strassler:2005qs}.\footnote{There is an apparent puzzle of the (ir)relevancy of the couplings: Classically $\lambda$ is irrelevant, while $\widetilde{\lambda}$ is relevant. However, for the duality to work, their (ir)relevancy must be the same. 
Indeed, due to the strong coupling effect, the (dual) quarks could receive significant anomalous dimensions in the IR. This resolves the puzzle: For example, in the $N_{f,1}=0$ case, $\lambda$ and $\widetilde{\lambda}$ are relevant when $N_{f,0}<2N_c$, marginal at $N_{f,0}=2N_c$, and irrelevant when $N_{f,0}>2N_c$.}
This lifts the modulus $\delta\Phi_{00}$, rendering the SUSY breaking vacua meta-stable \cite{Amariti:2006vk}.
The one-loop effective potential remains the same as the one without $W_{add}$, since $\delta\Phi_{00}$ does not couple to the SUSY breaking fields.

However, there is a notable difference in the SUSY vacua. Similarly to the previous case (\ref{NPplusmasslessPot}), The low energy effective potential at the energy scale $E<h\Phi_{00}, h\Phi_{11}$ yields
\be
W_L=N\left(h^{N_f}\Lambda^{3N-N_f}\det\,\Phi_{00}
\det\,\Phi_{11}\right)^{{1\over N}}
-h\mu^2\Tr\,\Phi_{11}+\widetilde{\lambda}\Tr\,\Phi_{00}^2\ .
\label{LEEP}
\ee
Integrating out $\Phi_{11}$, it becomes
\be
W_L=-(N_{f,1}-N)\left({\mu^{2N_{f,1}}\Lambda^{N_f-3N}}\over h^{N_{f,0}}\det\Phi_{00}\right)^{{1\over N_{f,1}-N}}
+\widetilde{\lambda}\Tr\,\Phi_{00}^2\ .\label{FUquartSUSYpot}
\ee
Thus the SUSY vacua appear at
\be
\Phi_{00}=\left({\mu^{2N_{f,1}}\Lambda^{N_f-3N}\over (-2\widetilde{\lambda})\, h^{N_{f,0}}}\right)^{{N_{f,1}-N\over N_f+N_{f,1}-2N}}{\bf 1}_{N_{f,0}}\ .
\ee
One can easily check that if $|\epsilon|=\left|{\mu\over\Lambda}\right|\ll 1$, the scales of SUSY breaking, SUSY vacua, and the Landau pole of the magnetic theory obey the inequalities
$\mu\ll h\Phi_{00}, h\Phi_{11}\ll\Lambda$.\footnote{Upon integrating out $\Phi_{11}$, we obtain $\Phi_{11}=\left(h^{N-N_f}\mu^{2N}\det\,\Phi_{00}/ \Lambda^{3N-N_f}\right)^{1/(N_{f,1}-N)}{\bf 1}_{N_{f,1}}$.}
This ensures the legitimacy of our analysis as in the ISS model.

Casero, Nu\~nez, and Paredes argued that their flavored MN geometry  in the IR is closely related to ${\cal N}=1$ $SU(N_c)$ SQCD with $N_f$ massless quarks and the quark quartic potential \cite{Casero:2006pt}. They provided several qualitative evidence for the claim.
Given that, it might be possible to see the dynamical SUSY breaking in meta-stable vacua by adding flavor D5-branes corresponding to massive quarks in their background \cite{Nunez:2003cf}: We take $N_c$ to be large and $N_{f,0}$, for example, to be $N_c-1$ which satisfies the rank condition. Then the number of massive flavors can be just a few, say, $N_{f,1}=3$ which is compatible with the IR free condition for the magnetic theory.\footnote{$N_{f,1}=2$ might as well be good, corresponding to $N=1$.} This justifies the probe approximation. 
However, one may worry that the quartic potential for the massive quarks too might be generated.
At this point, it is not entirely clear if that is the case. The massive flavor D5-branes have quite different embeddings from the massless flavor D5-branes that created the background \cite{Nunez:2003cf}.
So it could be that there is an embedding which corresponds to the massive quarks without the quartic potential. Then there is the hope that we might be able to see the meta-stable SUSY vacua in the probe approximation. 
However, even if that is the case, the SUSY vacua may not be visible in the probe approximation.
This is because the SUSY vacua involve the vev of mesons for the massless quarks. On the gravity side this may be accounted for only by a deformation of the background flavor branes.

In sum, the meta-stable SUSY breaking vacua might be realized in the flavored MN geometry with a few D5-brane probes, but the SUSY vacua may not.

\section{A model based on Klebanov-Strassler theory}

We next look for the models with dynamical SUSY breaking in meta-stable vacua which may be dual to the KS geometry with a few additional brane probes. 
The basic idea of finding meta-stable SUSY breaking vacua is the same as the previous model with the simple quartic potential.

The KW-KS gauge theory is ${\cal N}=1$ $SU(N_1=N+M)\times SU(N_2=N)$ gauge theory with massless bi-fundamentals and the quartic potential \cite{Klebanov:1998hh,Klebanov:2000hb}
\be
W=\lambda\Tr_a\det_{\alpha\dot{\alpha}}A_{\alpha i}B^i_{\dot{\alpha}}
=\lambda\left(A^a_{1i}B^i_{1b}A^b_{2j}B^i_{2a}-A^a_{1i}B^i_{2b}A^b_{2j}B^j_{1a}\right)\ ,
\label{KWpot}
\ee
where $A^a_{\alpha i}$ and $B^i_{\dot{\alpha} a}$ are bi-fundamentals $({\bf N_1}, \overline{{\bf N_2}})$ and $(\overline{{\bf N_1}}, {\bf N_2})$ respectively, and $\alpha, \dot{\alpha}=1,2$, $i=1,\cdots , N_1$, and $a=1,\cdots , N_2$.

We now add $N_f$ massive quarks ${\cal Q}_r=(Q_{ir}, Q'_{ar})$ and $\widetilde{{\cal Q}}^r=(\widetilde{Q}^{ri},\widetilde{Q'}^{ra})$ where $i=1,\cdots, N_1$, $a=1,\cdots, N_2$, and $r=1,\cdots, N_f$. They are (anti-)fundamentals of $SU(N_1)\times SU(N_2)$. The one-loop beta functions for the $SU(N_1)$ and $SU(N_2)$ gauge couplings are given by 
\bea
\beta^{(N_1)}&=&-{g_1^3\over 16\pi^2}(N+3M-N_f)\ ,\nn\\
\beta^{(N_2)}&=&-{g_2^3\over 16\pi^2}(N-2M-N_f)\ .\nn
\eea
We consider the case where the $SU(N_1)$ theory is asymptotically free, while the $SU(N_2)$ theory is IR free,
\be
N<2M+N_f\hspace{1cm}\mbox{and}\hspace{1cm}
N+3M>N_f\ .
\ee
We denote the strong coupling scale of the $SU(N_1)$ theory by $\Lambda_1$ and the Landau pole of the $SU(N_2)$ theory by $\Lambda_2$.

To study the IR physics of this theory, we dualize the $SU(N_1)$ theory to its magnetic description, that is, $SU(\widetilde{N}=N+N_f-M)$ SQCD with $2N+N_f$ flavors and $SU(N_1)$ singlets as well as the $SU(N_1)$ neutral components $Q'$ and $\widetilde{Q}'$ of the massive quarks.  
The superpotential of this magnetic dual theory takes the form (plus the mass term $\Tr\, mQ'\widetilde{Q}'$) \cite{Strassler:2005qs}
\be
W=h\left(\Tr\, {\cal A}^{\beta} Y_{\beta\dot{\beta}}{\cal B}^{\dot{\beta}}
+\Tr\, {\cal A}^{\beta} Z_{\beta}\widetilde{q}
+\Tr\, qZ_{\dot{\beta}}{\cal B}^{\dot{\beta}}\right)
+\left(h\Tr\, qZ\widetilde{q}
-h\mu^2\Tr\, Z\right)
+\widetilde{\lambda}\Tr\,\det_{\beta, \dot{\beta}} Y\ ,\label{KSdualspot}
\ee
where the trace is over $SU(\widetilde{N})$ and $SU(N_2=N)$ indices. 
The dual quarks ${\cal A}^{\beta}$ are $SU(\widetilde{N})$ fundamentals and ${\bf N_2}$, and 
the dual anti-quarks ${\cal B}^{\dot{\beta}}$ are $SU(\widetilde{N})$ anti-fundamentals and $\overline{{\bf N_2}}$.
Another dual quarks $q$ are $SU(\widetilde{N})$ fundamentals and $\overline{{\bf N_f}}$, and 
the dual anti-quarks $\widetilde{q}$ are $SU(\widetilde{N})$ anti-fundamentals and ${\bf N_f}$.
The $SU(N_1)$ singlets $Y_{\beta\dot{\beta}}$ are $(\overline{{\bf N_2}}, {\bf N_2})$, 
$Z_{\beta}$ $(\overline{{\bf N_2}}, {\bf N_f})$, $Z_{\dot{\beta}}$ $(\overline{{\bf N_f}}, {\bf N_2})$,
and $Z$ $({\bf N_f}, \overline{{\bf N_f}})$.
The singlets $Y_{\beta\dot{\beta}}$ correspond to the $AB$ mesons, $Z$ the $Q\widetilde{Q}$ mesons, $Z_{\beta}$ the $A\widetilde{Q}$ mesons, and $Z_{\dot{\beta}}$ the $QB$ mesons.

For the $SU(\widetilde{N})$ magnetic theory to be IR free, the theory has to be in the range
\be
3\widetilde{N}<2N+N_f\hspace{1cm}\Longleftrightarrow\hspace{1cm}N+2N_f<3M\ .
\ee
We denote the Landau pole of the $SU(\widetilde{N})$ theory by $\Lambda$.
Similarly to the ISS plus massless case with the superpotential (\ref{SUPOTISSplusmassless}), the SUSY breaking rank condition requires
\be
\widetilde{N}<N_f\hspace{.5cm}\Longleftrightarrow\hspace{.5cm}
N<M\ .
\ee
In order for the Seiberg dual description to exist, $\widetilde{N}$ must be greater than one. Hence the number $M$ is bounded by
\be
N<M<N+N_f-1\ .\label{conditiononM}
\ee

After dualizing the $SU(N_1)$ theory, the one-loop beta function for the $SU(N_2)$ theory becomes 
\be
\widetilde{\beta}^{(N_2)}=-{g_2^3\over 16\pi^2}\left(3N_2-2\widetilde{N}-N_f-2N_f-4N_2\right)
=-{g_2^3\over 16\pi^2}\left(2M-3N-5N_f\right)\ .
\ee
In addition to $2\widetilde{N}+N_f$ $SU(N_2)$ (anti-)fundamentals, the pair of mesons ($Z_{\beta},Z_{\dot{\beta}}$) are $2N_f$ quarks for the $SU(N_2)$ theory, and the mesons $Y_{\beta\dot{\beta}}$ are four adjoints. 
So if the energy scale we are probing at is around or above the mass scales of the $SU(N_2)$ matters but well below the new Landau pole $\widetilde{\Lambda}_2$,
in the range (\ref{conditiononM}) the one-loop beta function is positive and the $SU(N_2)$ theory remains IR free.

Since the $SU(N_2)$ adjoint matters $Y_{\beta\dot{\beta}}$ have the mass $\widetilde{\lambda}$, they are to be integrated out  below the energy scale $\widetilde{\lambda}$. So at this energy scale, the $SU(N_2)$ theory becomes asymptotically free.\footnote{The strong coupling scale $\Lambda'_2$ may be found by the matching of the couplings: $(\Lambda'_2)^{2M+N-5N_f}=\widetilde{\Lambda}_2^{2M-3N-5N_f}\widetilde{\lambda}^{4N}$, or equivalently $\Lambda'_2=\widetilde{\Lambda}_2(\widetilde{\lambda}/\widetilde{\Lambda}_2)^{4N/(2M+N-5N_f)}$. If $\widetilde{\lambda}>\widetilde{\Lambda}_2$, there is no reason to consider the IR free $SU(N_2)$ theory with the Landau pole $\widetilde{\Lambda}_2$ at all. We only consider the case $\widetilde{\lambda}\ll\widetilde{\Lambda}_2$.}
However, as we discuss below, we will be interested in the energy scale at or above $\mu$. Moreover, having the gauge/string duality in mind, we would like to consider the situation where the $SU(N_1)$ mesons $Y_{\beta\dot{\beta}}$ can be regarded as the light degrees of freedom. So we require that $\widetilde{\lambda}\sim\mu\, (\ll \widetilde{\Lambda}_2)$. Then the $SU(N_2)$ theory remains IR free around the energy scale $\mu$.

Under the conditions we discussed above, 
the dual $SU(\widetilde{N})\times SU(N_2=N)$ theory with the superpotential (\ref{KSdualspot}) (plus the mass term $\Tr\, mQ'\widetilde{Q}'$) is IR free, and the K\"ahler potential is smooth.
It is thus effective to use this dual description to study the IR physics. 
Similarly to the previous examples, the supersymmetry is spontaneously broken in all orders in perturbation theory, since the F-term
\be
F_{Z}={\del W\over\del Z}=h\left(\widetilde{q}q-\mu^2{\bf 1}_{N_f}\right)
\ee
cannot be vanishing due to the rank condition $\widetilde{N}<N_f$. 
The SUSY breaking vacua are at 
\bea
&&{\cal A}^{\beta}={\cal B}^{\dot{\beta}T}=0\ ,\hspace{.5cm} Z_{\beta}=({\bf 0}_{N\times \widetilde{N}}\ , \hat{Z}_{\beta})
\ ,\hspace{.5cm} Z_{\dot{\beta}}=\left(
\begin{array}{c}
{\bf 0}_{\widetilde{N}\times N} \\ 
\hat{Z}_{\dot{\beta}}
\end{array}
\right)\ ,\nn\\
&& q=\left(q_0\ , {\bf 0}_{\widetilde{N}\times (N_f-\widetilde{N})}\right)\ ,\hspace{.5cm}
\widetilde{q}^T=\left(\widetilde{q}_0\ , {\bf 0}_{\widetilde{N}\times (N_{f}-\widetilde{N})}\right)\ ,
\label{KSSUSYvacua}\\
&&Z=\left(
\begin{array}{cc}
{\bf 0}_{\widetilde{N}} & {\bf 0}_{\widetilde{N}\times (N_f-\widetilde{N})} \\
{\bf 0}_{(N_f-\widetilde{N})\times\widetilde{N}}  & \hat{Z}
\end{array}
\right)\ ,\hspace{.5cm}
Y_{\beta\dot{\beta}}=0\ ,\nn
\eea
where $q_0=\mu e^{\theta}{\bf 1}_{\widetilde{N}}$ and $\widetilde{q}_0=\mu e^{-\theta}{\bf 1}_{\widetilde{N}}$. The $N\times (N_f-\widetilde{N})$ matrices $\hat{Z}_{\beta}$ and $\hat{Z}_{\dot{\beta}}^{\dagger}$ must be equal due to the D-flatness condition $\Tr\left(\hat{Z}_{\beta}^{\dagger}T_A\hat{Z}_{\beta}-\hat{Z}_{\dot{\beta}}T_A\hat{Z}_{\dot{\beta}}^{\dagger}\right)=0$.
Note that $Y_{\beta\dot{\beta}}$, the analogue of $\Phi_{00}$ in the previous sections, is fixed to be zero due to the quadratic term $\widetilde{\lambda}\Tr\,\det_{\beta, \dot{\beta}} Y$ in the superpotential, the magnetic dual of the quartic potential. 
The vacuum energy density of the SUSY breaking vacua is $V_{min}=\left(N_f-\widetilde{N}\right)\left|h^2\mu^4\right|$.

The perturbative stability of the SUSY breaking vacua can be argued in much the same way as in the previous section: 
Let us consider the maximally symmetric vacua, $\hat{Z}=0$, $\theta=0$, $\hat{Z}_{\beta}=\hat{Z}_{\dot{\beta}}^{\dagger}=0$. The pseudo-moduli in this case are $\delta\hat{Z}$, $\mu\mbox{Re}\,\delta\theta$, and $\half\left(\delta\hat{Z}_{\beta}+\delta\hat{Z}_{\dot{\beta}}^{\dagger}\right)$.
The presence of the $SU(N_2)$ gauge fields is the only notable difference from the previous case.
However, the $SU(N_2)$ gauge fields do not directly couple to the SUSY breaking fields, the $\widetilde{N}\times (N_f-\widetilde{N})$ block of $q$ and $\widetilde{q}^T$, which are $SU(N_2)$ neutral. 
The $SU(N_2)$ matters acquire masses but all of order $\mu$, assuring the IR freedom and the weakness of the coupling around the energy scale $\mu$. 
So its presence will not affect the result for the one-loop effective potential.
Thus the pseudo-moduli are lifted exactly in the same way as in the previous case.

The SUSY vacua appear elsewhere in the field space of $Y_{\beta\dot{\beta}}$ and $Z$. 
At the energy scale $E<hY_{\beta\dot{\beta}}, hZ$, all the dual quarks can be integrated out.
The dual theory then becomes pure ${\cal N}=1$ $SU(\widetilde{N})$ Yang-Mills theory (plus $SU(\widetilde{N})$ neutral quarks $Q'$ and $\widetilde{Q'}$), ignoring the $SU(N_2)$ gauge fields. 
Similarly to (\ref{LEEP}), we find the low energy effective superpotential (plus the mass term $\Tr\, mQ'\widetilde{Q'}$)
\be
W_L=\widetilde{N}\left(h^{2N+N_f}\Lambda^{3\widetilde{N}-2N-N_f}\det_{SU(N_2)}\left(\det_{\beta,\dot{\beta}}Y\right)
\det\,Z\right)^{{1\over \widetilde{N}}}
-h\mu^2\Tr\,Z+\widetilde{\lambda}\Tr_{SU(N_2)}\det_{\beta,\dot{\beta}}Y\ .
\label{KSNPpot}
\ee
Integrating out $Z$ then yields
\be
W_L=-(N_f-\widetilde{N})\left({\mu^{2N_{f}}\Lambda^{2N+N_f-3\widetilde{N}}}\over h^{2N}\det_{SU(N_2)}\left(\det_{\beta,\dot{\beta}}Y\right)\right)^{{1\over N_{f}-\widetilde{N}}}
+\widetilde{\lambda}\Tr_{SU(N_2)}\det_{\beta,\dot{\beta}}Y\ .\label{KSSUSYpot}
\ee
Reinstating the $SU(N_2)$ gauge fields, this low energy effective theory is ${\cal N}=1$ $SU(N_2)$ gauge theory with the adjoint matters $Y_{\beta\dot{\beta}}$, $N_f$ massive quarks $Q'$, $\widetilde{Q'}$, and the superpotential (\ref{KSSUSYpot}).
The SUSY vacua are at 
\be
\det_{\beta,\dot{\beta}}Y=\left({\zeta^{{1\over M-N}}\over -\widetilde{\lambda}}\right)^{\frac{M-N}{M}}
{\bf 1}_N\ ,\label{SUSYvacuaY}
\ee
where $\zeta=\mu^{2N_{f}}\Lambda^{3M-N-2N_f}/h^{2N}$.
It is again easy to check that if $|\epsilon|=\left|{\mu\over\Lambda}\right|\ll 1$, the scale of SUSY breaking, the SUSY vacua, and the Landau pole of the dual theory obey the inequalities
$\mu\ll |hY_{\beta\dot{\beta}}|, |hZ|\ll\Lambda$. 
At this energy scale, the one-loop beta function is proportional to $3N_2-4N_2-N_f=-N_2-N_f$ and so is positive. 
The theory is IR free and in the weak coupling regime, and the $SU(N_2)$ gauge fields won't affect the low energy effective potential.
Assuming that $|\widetilde{\Lambda}_2|$ is not too smaller than $|\Lambda|$, this ensures the validity of our analysis.

Having the gauge/string duality in mind, we will be interested in fairly large $N$.
Then the minimal values of $M$ and $N_f$ which are compatible with all the above conditions are
\be
M=N+1\hspace{1cm}\mbox{and}\hspace{1cm}
N_f=3\ ,
\ee
corresponding to the $SU(2N+1)\times SU(N)$ KW-KS theory with $3$ massive quarks.
The case, $M=N+1$ and relatively small $N_f$, will be our favorite choice.

\medskip
We now consider the case when the $SU(N_1)$ theory is confining -- the number of flavors is equal to the number of colors plus one, that is,  $N_1+1=2N_2+N_f$, or equivalently $M+1=N+N_f$.
In this case the strongly coupled $SU(N_1)$ theory does not have the magnetic dual description. However, the non-perturbative superpotential is known in terms of the baryons and the mesons. 
The result is essentially the extrapolation of the $M<N+N_f-1$ case to $M=N+N_f-1$. 

We only discuss the minimal value case, $M=N+1$ and $N_f=2$. The gauge theory of our interest is thus ${\cal N}=1$ $SU(2N+1)\times SU(N)$ KW-KS theory with $2$ massive quarks.
For the $SU(N_1=2N+1)$ theory, the mesons and the baryons are given by 
\bea
\left({Y}_{\alpha\dot{\alpha}}\right)^a_b&=&A^a_{\alpha i}B^i_{\dot{\alpha} b}\ ,
\quad \left(Z_{\alpha}\right)^{as}=A^a_{\alpha i}\widetilde{Q}^{is}\ , 
\quad \left(Z_{\dot{\alpha}}\right)_{r b}=Q_{ir}B^i_{\dot{\alpha}b}\ , 
\quad Z_r^s=Q_{i r}\widetilde{Q}^{i s}, \nn\\
q_r&=&\epsilon^{i_1\cdots i_{N_1}}\left(A^{a_1}_{\alpha_1}\right)_{i_1}\cdots \left(A^{a_{N_1-1}}_{\alpha_{N_1-1}}\right)_{i_{N_1-1}}Q_{i_{N_1}r}\ , \nn\\
\widetilde{q}^s&=&\epsilon_{i_1\cdots i_{N_1}}\left(B_{\dot{\alpha}_1 a_1}\right)^{i_1}\cdots \left(B_{\dot{\alpha}_{N_1-1} a_{N_1-1}}\right)^{i_{N_1-1}}\widetilde{Q}^{i_{N_1}s}\ ,\\ 
{\cal A}^{\alpha_{N_1-1}}_{a_{N_1-1}}&=&\epsilon_{(a_1, \alpha_1)\cdots (a_{N_1-1}, \alpha_{N_1-1})}\epsilon^{i_1\cdots i_{N_1}}\left(A^{a_1}_{\alpha_1}\right)_{i_1}\cdots \left(A^{a_{N_1-2}}_{\alpha_{N_1-2}}\right)_{i_{N_1-2}}Q_{i_{N_1-1} r}Q_{i_{N_1} s}\ ,\nn\\ 
{\cal B}^{\dot{\alpha}_{N_1-1} a_{N_1-1} }&=&\epsilon^{(\dot{\alpha}_1,a_1)\cdots (\dot{\alpha}_{N_1-1},a_{N_1-1})}\epsilon_{i_1\cdots i_{N_1}}\left(B_{\dot{\alpha}_1 a_1}\right)^{i_1}\cdots \left(B_{\dot{\alpha}_{N_1-2}a_{N_1-2}}\right)^{i_{N_1-2}}\widetilde{Q}^{i_{N_1-1} a_{N_1-1}}\widetilde{Q}^{i_{N_1} a_{N_1}}\ , 
\nn
\eea
where $\alpha, \dot{\alpha}=1,2$, $i=1,\cdots, 2N+1$ for color $N_1$, $a,b=1,\cdots, N$ for color $N_2$, and $r,s=1,2$ for flavor $N_f$. There are not enough non-vanishing components of the quarks to compose any other baryons (after the gauge and global symmetry rotations).
The baryons can be heuristically thought of as the dual quarks for the \lq\lq $SU(\widetilde{N}=1)$" theory, and the mesons as \lq\lq $SU(\widetilde{N}=1)$" singlets.

The superpotential is essentially the sum of the perturbative potential (\ref{KSdualspot}) and the non-perturbative potential (\ref{KSNPpot}) with $\widetilde{N}=1$ and the appropriate normalization of the fields similar to (\ref{confsuppot}).
Hence this case also exhibits the dynamical SUSY breaking in meta-stable vacua.

In summary under certain conditions the KW-KS gauge theory with light quarks has meta-stable SUSY breaking vacua in the deeper IR in addition to the SUSY vacua. As an example, the $SU(2N+1)\times SU(N)$ KW-KS theory with $N_f\ge 2$ quarks and $N\gg 1$ satisfies the requisite conditions.

\section{Discussions}

The KS geometry is dual to the ${\bf Z}_2$ symmetric baryonic branch of the $SU((k+1)N)\times SU(kN)$ KW-KS theory which cascades to the $SU(2N)\times SU(N)$ theory in the IR \cite{Aharony:2000pp,Gubser:2004qj,Dymarsky:2005xt,Benna:2006ib}.
Thus we propose that the meta-stable SUSY breaking vacua might be visible in the KS geometry with one fractional D3/wrapped D5-brane and a few D7-branes as probes:
Let us consider the $SU(N_1)$ confining case, $M=N+1$ and $N_f=2$. At the SUSY breaking vacua, the baryons $q=\widetilde{q}^T\ne 0$ for which $U(1)_B$ is broken and there is a ${\bf Z}_2$ symmetry $q\leftrightarrow\widetilde{q}^T$. 
Note in particular that we can choose
\be
\left|\epsilon^{i_1\cdots i_{N_1}}\left(A^{a_1}_{\alpha_1}\right)_{i_1}\cdots \left(A^{a_{N_1-1}}_{\alpha_{N_1-1}}\right)_{i_{N_1-1}}\right|=\left|\epsilon_{i_1\cdots i_{N_1}}\left(B_{\dot{\alpha}_1 a_1}\right)^{i_1}\cdots \left(B_{\dot{\alpha}_{N_1-1} a_{N_1-1}}\right)^{i_{N_1-1}}\right|\ne 0\ ,
\label{Z2baryonvev}
\ee
and the mesons $Y_{\alpha\dot{\alpha}}=A_{\alpha}B_{\dot{\alpha}}=0$.
So the meta-stable SUSY breaking vacua can be built on the ${\bf Z}_2$ symmetric baryonic branch of the $SU(2N)\times SU(N)$ theory.
However, the SUSY vacua may not fit in the baryonic branch. Obviously the $U(1)_B$ symmetry is not broken at the SUSY vacua. Moreover, the mesons $Y_{\alpha\dot{\alpha}}$ acquire the vevs which would be interpreted as D3-branes in the bulk of deformed conifold, while the baryonic branch corresponds to a BPS bound state at threshold of $2N$ wrapped D5 and $N$ anti D5-branes at the tip \cite{Dymarsky:2005xt}.
This seems to suggest that the meta-stable SUSY breaking vacua might be realized in the KS geometry by adding a few probe branes, but the SUSY vacua may not.
The decay from the SUSY breaking to SUSY vacua may correspond to the process $N$ D5-branes paired up with $N$ anti D5-branes to form $N$ D3-branes escaping to the bulk and absorbed into the D7-branes off the tip. 
So the KS geometry has to be deformed in the decay process.

However, among other things, we have a few points to worry about. It was necessary to consider $|\widetilde{\lambda}|\sim |\mu|\ll |\Lambda|, |\widetilde{\Lambda}_2|$ and $|\widetilde{\Lambda}_2|$ not too smaller than $|\Lambda|$, for our analysis to be trustable.
It is not clear if these conditions can be met when $N$ and $g_sN$ ($g_s\ll 1$) are large on the gravity side.
If not, we might have to go beyond the classical supergravity approximation in order to see the meta-stable SUSY vacua.
Also as we take $N$ large, $|Y_{\beta\dot{\beta}}|$ at the SUSY vacua comes close to $|\Lambda|$, as we can see from (\ref{SUSYvacuaY}). This renders our field theory analysis of SUSY vacua unreliable, although the SUSY vacua may not be seen in the supergravity probe approximation in any case.

Nonetheless let us remark on the probe D7-branes in the KS geometry.
First, the supersymmetric embeddings of probe D7-branes in the KS geometry dual to massive quarks were studied in \cite{Karch:2002sh,Ouyang:2003df,Kuperstein:2004hy}. In addition to the mass term, these embeddings typically generate additional quartic terms of the type ${\cal Q}AB\widetilde{{\cal Q}}$ \cite{Ouyang:2003df,Kuperstein:2004hy} which translate to $Q'Y\widetilde{Q'}$ and $Z_{\beta}Z_{\dot{\beta}}$. 
So in discussing the gauge/string duality these types of terms should be included in our dual gauge theory analysis. However, their presence does not seem to afflict or alter much of our analysis. 
We suspect that these supersymmetric embeddings correspond to the vacua (\ref{KSSUSYvacua}) in the $M=N$ (or equivalently $\widetilde{N}=N_f$) case where the F-term $F_Z$ vanishes and the SUSY is not broken. 
Second, there exists in fact a perturbatively stable non-supersymmetric embedding of D7-branes in the KS geometry \cite{Sakai:2003wu}. Their embedding requires nontrivial gauge fields and  necessarily induces extra D3-brane charge on the D7-branes. This is perhaps the type of embeddings we are after, although their embedding does not have the free parameter which might have corresponded to the mass scale $\mu$.
It might be that their D7-branes or similar can support an instanton-like gauge configuration localized near the tip which represents a wrapped D5-brane and whose size corresponds to the mass scale $\mu\sim\langle q\rangle$.



Finally, provided that our claim be confirmed, we may use this SUSY breaking mechanism in the KS throat of the GKP-KKLT flux compactification \cite{Giddings:2001yu} \cite{Kachru:2003aw}. It is somewhat in a similar spirit as the models considered in \cite{Dudas:2006gr,Abe:2006xp,Kallosh:2006dv}, but the ISS-like SUSY breaking is now realized entirely in the gravitational sector as in \cite{Argurio:2006ny,Kachru:2002gs}.
This would provide a new natural brane setup to the uplifting of the vacuum energy leading to de Sitter vacua. It would be very interesting to explore this possibility further.

\section*{Acknowledgements}

I would like to thank Adel Awad, Sumit Das, Paolo Di Vecchia, Mads Frandsen, Masafumi Fukuma,  Niels Obers, Thomas Ryttov, Al Shapere,  especially Philip Argyres and Aki Hashimoto for helpful discussions and comments. I am especially grateful to the String Theory Group at University of Kentucky for their great hospitality and providing me invaluable time. This work was supported in part by the European Community's Human Potential Programme under contract MRTN-CT-2004-005104 \lq Constituents, fundamental forces and symmetries of the universe'.


\end{document}